\documentclass[journal=aamick,manuscript=article,layout=traditional]{achemso}
\setkeys{acs}{maxauthors=6,etalmode=truncate}

\usepackage{natbib}
\usepackage{graphicx}

\usepackage{color,soul}

\title{On the effect of surface functional groups on the structural and dynamical properties of an ionic liquid confined in zeolite templated carbons studied through molecular simulations}

\author{El Hassane Lahrar}
\affiliation[1]{CIRIMAT, Universit\'e de Toulouse, CNRS, B\^at. CIRIMAT, 118, route de Narbonne 31062 Toulouse cedex 9, France}
\alsoaffiliation[2]{R\'eseau sur le Stockage \'Electrochimique de l'\'Energie (RS2E), F\'ed\'eration de Recherche CNRS 3459, HUB de l'\'Energie, Rue Baudelocque,  80039 Amiens, France}

\author{Irena Deroche}
\affiliation[3]{Universit\'e de Haute-Alsace, Institut de Science des Mat\'eriaux de Mulhouse (IS2M), CNRS UMR 7361, F-68100 Mulhouse, France}
\alsoaffiliation[4]{Universit\'e de Strasbourg, F-67081 Strasbourg, France}

\author{Cam\'elia Matei Ghimbeu}
\affiliation[3]{Universit\'e de Haute-Alsace, Institut de Science des Mat\'eriaux de Mulhouse (IS2M), CNRS UMR 7361, F-68100 Mulhouse, France}
\alsoaffiliation[4]{Universit\'e de Strasbourg, F-67081 Strasbourg, France}
\alsoaffiliation[2]{R\'eseau sur le Stockage \'Electrochimique de l'\'Energie (RS2E), F\'ed\'eration de Recherche CNRS 3459, HUB de l'\'Energie, Rue Baudelocque,  80039 Amiens, France}

\author{Patrice Simon}
\affiliation[1]{CIRIMAT, Universit\'e de Toulouse, CNRS, B\^at. CIRIMAT, 118, route de Narbonne 31062 Toulouse cedex 9, France}
\alsoaffiliation[2]{R\'eseau sur le Stockage \'Electrochimique de l'\'Energie (RS2E), F\'ed\'eration de Recherche CNRS 3459, HUB de l'\'Energie, Rue Baudelocque,  80039 Amiens, France}

\author{C\'eline Merlet}
\affiliation[1]{CIRIMAT, Universit\'e de Toulouse, CNRS, B\^at. CIRIMAT, 118, route de Narbonne 31062 Toulouse cedex 9, France}
\alsoaffiliation[2]{R\'eseau sur le Stockage \'Electrochimique de l'\'Energie (RS2E), F\'ed\'eration de Recherche CNRS 3459, HUB de l'\'Energie, Rue Baudelocque,  80039 Amiens, France}
\phone{+33 (0)5 61 55 62 84}
\email{merlet@chimie.ups-tlse.fr}

\date{}


\begin{document}


\begin{abstract}
Porous carbons are used in a wide range of applications, including electrochemical double layer capacitors for energy storage, in which electrolyte ion properties under confinement are crucial for the performance of the systems. While many synthesis techniques lead to the presence of surface functional groups, their effect on the adsorption and diffusion of electrolyte ions is still poorly understood. In this study we investigate the effect of surface chemistry on dynamical and structural properties of adsorbed ions through molecular dynamics simulations of a neat ionic liquid in contact with several zeolite templated carbons. The steric and specific influence of functional groups is explored using three structures: a structure without any functional groups; a structure with ester, hydroxyl, anhydride acid and carboxyl functional groups; and a structure where the oxygen and hydrogen atoms are replaced by carbon atoms. We show that the functionalization has a limited impact on the structure of the confined electrolyte but affects the diffusion coefficients, and that the relative importance of the structure and chemical nature of the functional groups is not the same depending on the ion type and property considered.
\end{abstract}

\section{Introduction}

Electrochemical double layer capacitors, also known as supercapacitors, are characterized by large power densities, long cycle lifetimes and a wide range of operating temperatures.~\citep{Zhong15,Beguin14,Brandt13,Lewandowski10} In these systems, the charge storage is based on the voltage-driven electrosorption of ions from an electrolyte onto the electrode surfaces~\citep{Simon-Gogotsi,Forse16,Salanne16}. The electrodes commonly consist in porous carbons thanks to the large surface area, good electronic conductivity, good stability and ease of synthesis of these materials~\citep{Simon13}. Numerous studies have been carried out to understand what parameters affect the capacitance and, as a consequence, the energy stored in these systems. Important features include the pore size and how similar it is to the ion size, and the electrode curvature~\citep{Feng13,Chmiola06,Raymundo-Pinero06,Merlet12, Huang-curv,Merlet13d,Feng-curv}. While many synthesis techniques, such as activation at high temperature with an acid,~\citep{Oliveira-activ,Kim-activ} lead to surface functional groups, the effect of their presence on ion electrosorption  and dynamics of the confined electrolyte have not been widely studied. The existing literature demonstrates that the surface composition (presence of oxygen, nitrogen, ...) impacts significantly the density of the confined electrolyte and its mobility.~\citep{Sint08,Dyatkin14,Dyatkin15,Dyatkin16,Dyatkin18,Kerisit14} This is not surprising as the surface chemistry influences many properties including the electronic conductivity, pore accessibility, wettability, and polarity.

A number of experimental studies on the effect of functionalization on supercapacitors performance and properties of the confined electrolyte have used idealized structures such as graphene nanoplatelets (GNPs) and carbon nano-onions (CNOs), or Carbide-Derived Carbons (CDCs) as electrodes.~\citep{Moussa16,Dyatkin14,Dyatkin15,Dyatkin16,Dyatkin18} Moussa~\emph{et~al.} have employed CNOs to investigate the relationship between surface chemistry and defects, and the capacitance in aqueous and organic electrode.~\citep{Moussa16} They have systematically controlled the surface chemistry and amount of active sites by thermal oxidation under air and thermal reduction under hydrogen while keeping the textural properties, characterized by the specific surface area, similar. They have shown an increase of capacitance followed by a plateau when the amount of oxygen groups increases, in opposition to a linear increase with the amount of active sites in the case of defects. This demonstrates the importance of the active sites, and active surface area, for the determination of the capacitance.

Realising vacuum annealing at different temperatures, it is possible to synthesize CDCs with similar pore sizes and pore volumes but different surface chemistries and crystallinities. In particular, vacuum annealing at higher temperatures, increases ordering, reduces the number of defects, and leads to higher electronic conductivities. Using such an approach, Dyatkin~\emph{et~al.}~\citep{Dyatkin14} have studied the electrochemical properties of carbon-based supercapacitors with 1-ethyl-3-methylimidazolium bis(trifluoromethylsulfonyl)imide, [EMI][TFSI], as an electrolyte. They have shown that functionalization does not necessarily affect the capacitance negatively even though functionalized carbons are usually associated with lower electronic conductivities compared to carbons vacuum annealed at high temperatures. However, graphitization indeed tends to improve the performance at high rates. In later works, Dyatkin~\emph{et~al.}~\citep{Dyatkin15,Dyatkin16} have demonstrated using quasi-elastic neutron scattering and inelastic neutron scattering that the nature of the functional groups is very important. In particular, hydrogenated and oxygenated CDCs lead to faster ion diffusion and stronger ion-electrode interactions while aminated CDCs induce reduced performance compared to both hydrogenated/oxygenated and defunctionalized CDCs. The oxygen and hydrogen functional groups modify the ionophilicity of the surface and draw the ions closer to the carbon surface rendering the cations more mobile. In non-functionalized carbons, the ion packing is denser which reduces the ion mobility. In these works, the experimental results are completed and supported by molecular dynamics simulations (MD) of the same electrolyte in carbon slit pores. In many cases, the functional groups lead to pseudocapacitive phenomena which can also enhance the performance of supercapacitors.~\citep{Beguin14} 

Molecular simulations on different systems have shown that it is not only the nature of the functional groups that matters but actually the electrode-electrolyte combination. Kerisit~\emph{et~al.}~\citep{Kerisit14} have conducted MD simulations of 1-butyl-3-methylimidazolium trifluoromethanesulfonate, [BMIM][OTf], in contact with graphene and varying the amount of oxygen-containing functional groups (hydroxyl, epoxy). They have shown that the oxygen containing functional groups affect capacitive properties negatively. In this case, there is a strong hydrogen bonding between the surface groups and SO$_3$, which restrains the reorientation and mobility of the anions, leading to a decrease in capacitance. Other MD simulations studies of ionic liquids based on the tetrafluoroborate anion ([EMIM][BF$_4$], [BMIM][BF$_4$]) in contact with planar graphene oxide and graphene oxide nanochannels have also shown a negative impact on the ion dynamics.~\citep{DeYoung14,Wang18} More recently, Schweiser~\emph{et al.}~\citep{Schweiser19} have used DFT calculations to study a range of surface chemistries and electrolytes, including the effect of solvation. This DFT study, combined with some electrochemical experiments suggests that interaction energies provide a good descriptor for the prediction of the capacitance. In all these theoretical studies, the electrodes are considered through very simplified geometries (slit pores, planar surfaces) and the addition of functional groups is done randomly or in a regular fashion but not in a realistic way. This limits the understanding of ion organisation and diffusion in three-dimensional porous networks.

In this work, we investigated the effect of the presence of surface functional groups on the structural and dynamical properties of a neat ionic liquid in contact with a Zeolite Templated Carbon (ZTC). ZTCs are inorganic highly ordered microporous carbons with many attractive features~\citep{Nishihara18,Taylor20,Builes11,Roussel07, Beauvais05}. They are characterized by a large specific area, a high porosity, a very narrow and ordered pore size distribution, surface hydrophobicity and robustness. These materials have been tested for several applications such as electrochemical supercapacitors~\citep{Jurewicz}, lithium intercalation~\citep{Meyers} or hydrogen adsorption storage~\citep{Gadiou}. We carried out simulations on three different ZTC structures: i) an initial one without any functional groups,~\citep{Roussel07} ii)  a second one derived from the first, with ester, hydroxyl, anhydride acid and carboxyl functional groups, and iii) a third one where oxygen and hydrogen atoms are replaced by carbon atoms to be able to distinguish between steric and specific effects of the functional groups. The molecular simulations allowed us to calculate a number of structural and dynamical properties to see the effect of functional groups on the confined ionic liquid.

\section{Systems studied and methods}

\subsection{Systems studied}

\begin{figure}[ht!]
\centering
\includegraphics[scale=0.58]{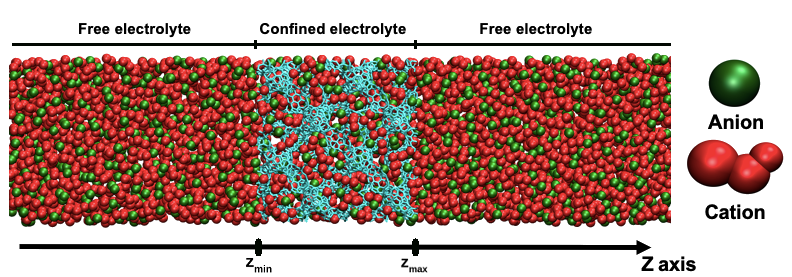}
\caption{Snapshot of one of the simulated systems consisting in a pure ionic liquid in contact with a porous carbon. Anions are represented in green, cations in red and carbon atoms in light blue. This snapshot was generated using the VMD software.\citep{VMD}}
\label{fig:setup}
\end{figure}

In this study we carried out molecular dynamics simulations of a neat ionic liquid, 1-butyl-3-methylimidazolium hexafluorophosphate ([BMIM][PF$_6$]), in contact with a ZTC as represented in Figure~\ref{fig:setup}. The ionic liquid is described by a coarse-grained model consisting of one site for the anion and three sites for the cation~\citep{Roy10b}. The ions are considered non polarisable. The intermolecular interactions are calculated as the sum of Lennard-Jones and coulombic interactions: 
\begin{equation}
u_{ij}(r_{ij})=4\varepsilon_{ij} \left [ \left ( \frac{\sigma_{ij}}{r_{ij}}\right )^{12}- \left ( \frac{\sigma_{ij}}{r_{ij}}\right )^6 \right ]+\frac{q_iq_j}{4\pi\varepsilon_0r_{ij}}
\end{equation}
where $r_{ij}$ is the distance between sites $i$ and~$j$, $\sigma_{ij}$ and $\epsilon_{ij}$ are the Lennard-Jones parameters defining respectively the position of the repulsive wall and the depth of the energy well, $q_i$ is the charge of site $i$, and $\varepsilon_0$ is the permittivity of free space. The ions have a charge of $\pm0.78$e. Reducing the charge from $\pm1.0$e to $\pm0.78$e allows for a remarkable agreement between simulation and experiment, or simulation with an all-atom model,~\citep{Canongia-Lopes04} for a variety of static and dynamic properties of the ionic liquid in the bulk,~\citep{Roy10a,Roy10b} at the interface with a planar electrode~\citep{Merlet11} and confined in a porous carbon.~\citep{Lahrar20} 

\begin{figure}[ht!]
\centering
\includegraphics[scale=0.24]{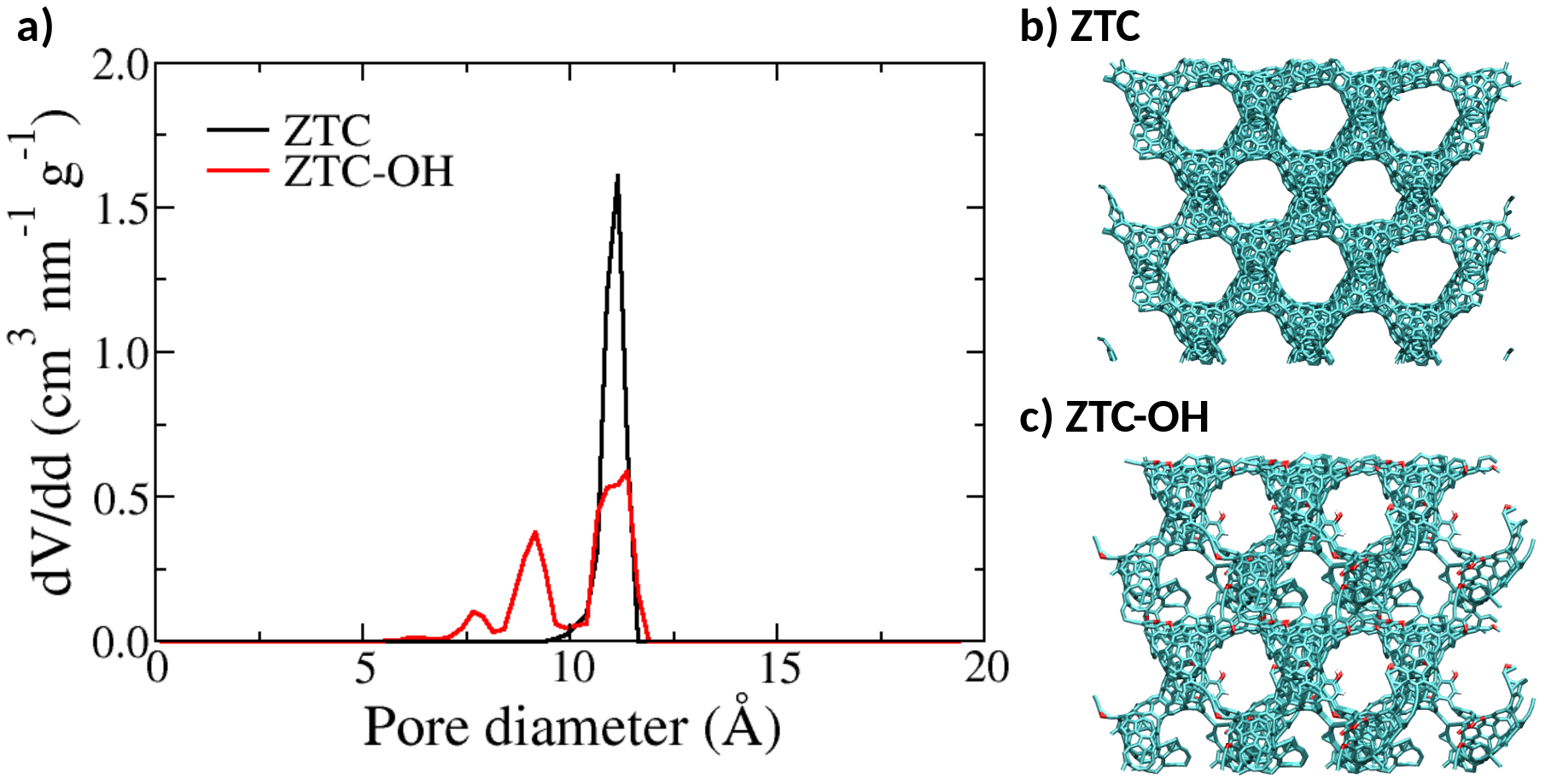}
\caption{Pore size distributions obtained using Poreblazer~\citep{Sarkisov11} (a) and atomic structures for non functionalized (b) and functionalized (c) porous carbons.}
\label{fig:PSD-FAU}
\end{figure}

We conducted simulations with three model carbon structures:\\
- the first structure, designated as ZTC, contains only carbon atoms and no functional groups;~\citep{Roussel07}\\
- the second structure, derived from the first and designated as ZTC-OH, contains carbon atoms and, oxygen and hydrogen functional groups;~\footnote{Ghimbeu, C and Sonnet, Ph. and Deroche, I., Realistic microscopic model of the FAU-type zeolite templated carbon, 2021,
in preparation}
- the third structure, designated as ZTC-C, is the same as ZTC-OH but the oxygen and hydrogen atoms have been replaced by carbon atoms.\\
The ZTC structure, generated by Roussel~\emph{et~al.}~\citep{Roussel07} using Grand Canonical Monte Carlo simulations, is a fully carbonaceous perfect negative of the templating zeolite (NaY – sodium exchanged Faujasite). It is characterized by a narrow unimodal pore size distribution as shown in Figure~\ref{fig:PSD-FAU}. Therefore, it differs from the synthesized ZTC by absence of both physical and chemical defects. The ZTC-OH model was conceived in order to reproduce physico-chemical and chemical properties of the zeolite replica through a two stage process, introducing (1) physical defects and (2) heterogeneous surface groups. First, a model reproducing the ZTC experimental structural properties such as porous volume, specific surface and density~\citep{Roussel07} was achieved through a combination of a Monte Carlo based simulation of acid leaching followed by geometry optimization. Next, heterogeneous surface functions have been introduced within the model, according to the experimental ZTC elemental composition.~\citep{Nishihara09} The heteroatoms form the following functional groups: ester, phenolic hydroxyl, anhydride acid and carboxyl, respectively with ratios of 7:2:2:1. Finally, the model was geometry optimized using a periodic Density Functional Theory (DFT) calculation.~\footnote{Ghimbeu, C and Sonnet, Ph. and Deroche, I., Realistic microscopic model of the FAU-type zeolite templated carbon, 2021,
in preparation}

As a consequence of the carbon modification procedure, the pore size distribution of ZTC-OH is quite different from the one of ZTC. It is largely due to the added functional groups, 128 oxygen and 32 hydrogen atoms, occupying the main pores and therefore generating smaller pores. The force field parameters for carbon, oxygen and hydrogen atoms are provided in Table~\ref{tableau:FAU}.~\citep{Cole83,Jorgensen96} The carbon atoms of the ZTC and ZTC-C structures are always neutral, while for ZTC-OH, the charges on carbon, oxygen and hydrogen atoms correspond to the charges obtained by DFT calculations with very small modifications to obtain a neutral material. The atomic charges of the ZTC-OH structure are available in the relevant input files in the Zenodo repository with identifier 10.5281/zenodo.4446690.

We note that, in this work, the carbon structure is considered as a rigid entity while in reality the structure is flexible to some extent.~\citep{Hantel11,Hantel12,Hantel14} Some rare theoretical studies on water-carbon systems show that the flexibility has a limited impact on properties of water~\citep{Werder03,Thomas09,Alexiadis08}. For systems similar to the ones of interest here, we have shown in a previous study that a flexibility leading to relative height changes of 1\% and 2\% does not affect dynamical and structural properties significantly.~\citep{Lahrar20}   

\begin{table}[ht!]
\centering
\begin{tabular}{|c|c|c|c|c|c|}
\hline
Atom & M (g~mol$^{-1}$) & $\sigma_i$ (\r{A}) & $\varepsilon_i$ (kJ~mol$^{-1}$) & $q_i$ (e) \\
\hline
C & 12.01 & 3.37 & 0.23 & 0.0 in ZTC/ZTC-C, \\
 & & & & heterogeneous in ZTC-OH\\
\hline
O & 15.99 & 3.16 & 0.037 & heterogeneous \\
 \hline
H & 1.00 & 0.00 & 0.00 & heterogeneous\\
 \hline
\end{tabular}
\caption{Force field parameters used for carbon, oxygen and hydrogen atoms.~\citep{Cole83,Jorgensen96}}
\label{tableau:FAU}
\end{table}

All simulations reported here have been conducted using LAMMPS~\citep{LAMMPS} and the initial configurations were generated using the fftool software.~\footnote{fftool, Ag\'ilio A. H. P\'adua, http://doi.org/10.5281/zenodo.18618} The rigidity of the cation is maintained using the SHAKE algorithm.~\citep{Ryckaert} The timestep is set to 2~fs. The simulations contain 1200 ion pairs and the simulated systems have dimensions close to 49 \r{A} $\times$ 49 \r{A} $\times$ 225~\r{A} depending on the carbon structure considered. The systems were first equilibrated in the NPT ensemble for 2~ns before collecting data for 10~ns in the NVT ensemble. The pressure of the NPT simulations is set to 1~atm and all simulations are done at 400~K. The barostat and thermostat time constants are 0.5~ps and 0.1~ps respectively. A cut-off of 12~\r{A} was used for the Lennard-Jones interactions while coulombic interactions were evaluated using a particle-particle particle-mesh Ewald solver. 

\subsection{Pair distribution functions}

The local structure of a system of particles (atoms, molecules, colloids) can be determined from the calculation of pair distribution functions (or radial distribution functions), denoted by $g(r)$. These functions correspond to the probability of finding a pair of particles at a distance $r$, with respect to the estimated probability for a completely random distribution at the same density. In other words, they describe the variation in density as a function of the distance from a reference particle. A possible expression for these pair distribution functions is:
 \begin{equation}
  g_{\alpha \beta}(||\mathbf{r_i}-\mathbf{r_j}||) = \frac{\rho_{\alpha\beta}^{(2)}(\mathbf{r_i},\mathbf{r_j})}{\rho_{\alpha}^{(1)}(\mathbf{r_i})\times\rho_{\beta}^{(1)}(\mathbf{r_j})},
 \end{equation}
where $\rho_{\alpha}^{(1)}$ and $\rho_{\alpha \beta}^{(2)}$ are respectively the one-body and two-body particle densities for ions of species $\alpha$ (and $\beta$). The pair distribution functions allow for the comparison of environments around an ion in different electrolytes and also to characterize the structural changes associated with confinement.

\subsection{Diffusion coefficients}

Diffusion coefficients are very often calculated to characterize the dynamic behavior of a fluid. For a uniform fluid, where the density is the same at any point in space, the diffusion is homogeneous and equal along the three axes $x$, $y$ and $z$. The determination of the self-diffusion coefficients is done using Einstein's relation which relates this property to the mean-square displacement of the molecules:
 \begin{equation}
  D = \lim_{t\to\infty} \frac{1}{2dt} <| \Delta \mathbf{r_i}(t) |^2>,
 \end{equation}
where $d$ is the dimensionality of the system and $\Delta \mathbf{r_i}(t)$ is the displacement of a typical ion of the considered species in time $t$.

The presence of the porous carbon breaks the symmetry of the system, as shown in Fi\-gu\-re~\ref{fig:setup}, and two regions corresponding to the ``free electrolyte" and the ``confined electrolyte" can be defined. The determination of diffusion coefficients in such a system has been described in the literature~\citep{Liu04,Rotenberg07, Lahrar20}. An analysis in which fictitious boundaries are introduced is used. $D_{xx}$ and $D_{yy}$ are determined from the mean square displacements $<\Delta x^2(t)>$ and $<\Delta y^2(t)>$ of particles remaining in a given region. $P_i(t)$ is the survival probability for a particle in that given region :  
\begin{equation}
    D_{xx}(z_i) =  \lim\limits_{t \rightarrow \infty} \frac {<\Delta x_i^2(t)>} {2tP_i(t)},\hspace{0.5cm}   
    D_{yy}(z_i) =  \lim\limits_{t \rightarrow \infty} \frac {<\Delta y_i^2(t)>} {2tP_i(t)}
\end{equation}
$P_{i}(t)$ is the probability for a particle $i$ to remain in a region of interest and can be calculated numerically from the simulation. Let $N(t,t+\tau)$ be the total number of $i$ particles in the region of interest during the time interval between $t$ and $t+\tau$, and $N(t)$ designate the number of particles in the layer at time $t$. Then $P(\tau)$ can be calculated as: 

\begin{equation}
  P(\tau) = \frac {1}{T}\sum_{t=1}^{T}\frac{N(t,t+\tau)}{N(t)}
 \end{equation}
 where $T$ is the total number of time steps averaged over.
 
The diffusion coefficient along the $z$ axis, $D_{zz}$, is determined from the autocorrelation of an eigenfunction based on the $z$ limit condition : 
\begin{equation}
D_{zz}(z_i)=-\left(\frac{L}{n\pi}\right)^2  \lim\limits_{t \rightarrow 0} \frac {\ln(<\psi_n^i(t) \psi_n^i(0)>)} {t},
\end{equation}
where $\psi_n^i(t)$ is given by:
\begin{equation}
\psi_n^i(z_i)= \sin\left(n\pi \frac{z(t)-z_{min}^i}{z_{max}^i-z_{min}^i}\right).
\end{equation}
In these equations, $z_{min}^i$ and $z_{max}^i$ are the coordinates that define the width of a given region $L = z_{max}^i - z_{min}^i $, and $n$ is an integer which should not affect the results much in the diffusive regime, here we choose $n=3$. This result is based on the Smoluchowski equation applied in a region where the potential of mean force is constant. More details on this analysis are available in reference~\citep{Liu04}.

\subsection{Adopted strategy}

As mentioned earlier, the addition of heteroatoms corresponding to functional groups (128 oxygens and 32 hydrogens) in the ZTC structure changes the pore size distribution significantly. 
\begin{figure} [ht!]
\begin{center}
\includegraphics[scale=0.25]{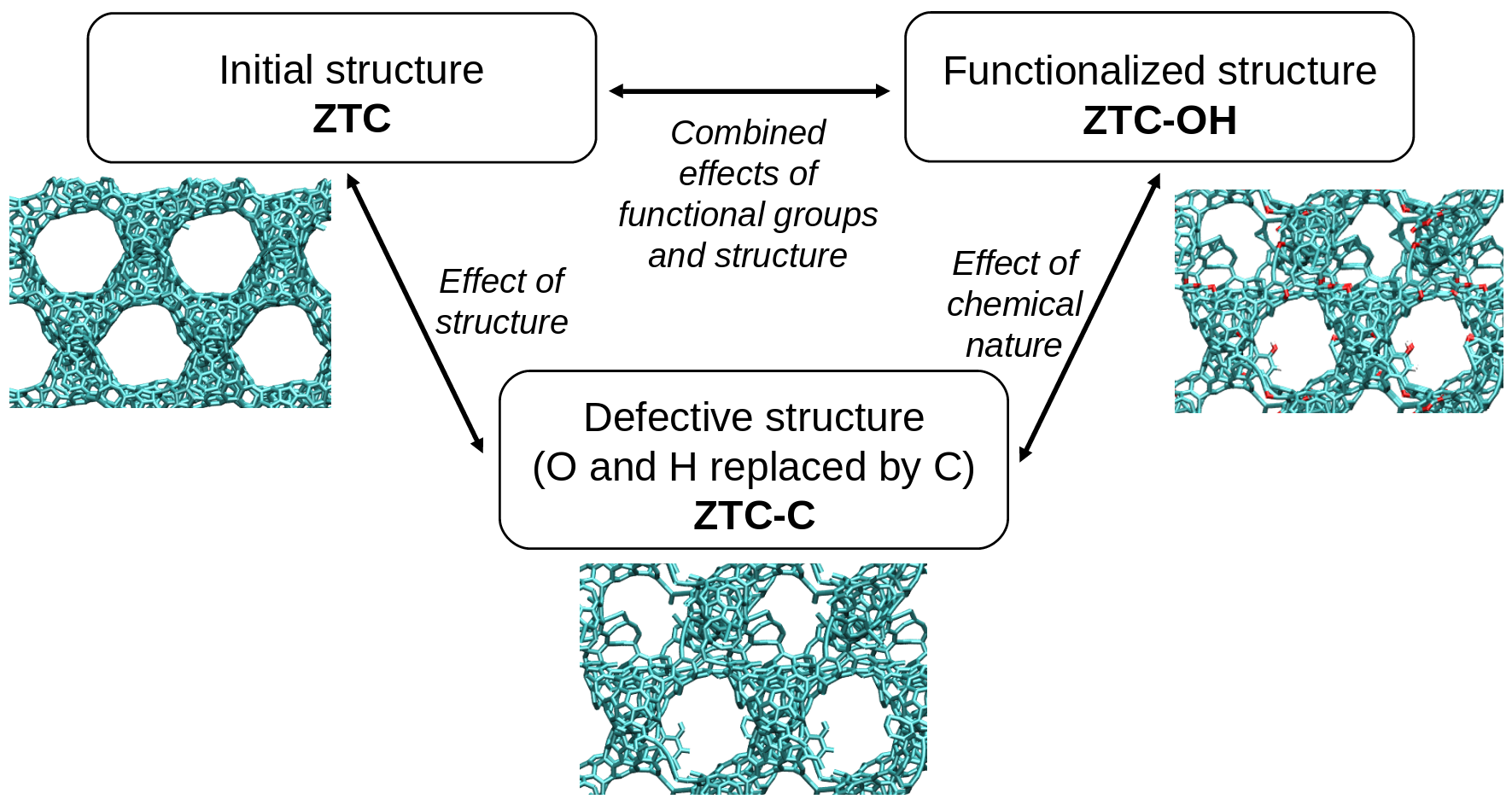}
\end{center}
\caption{Strategy used for the different simulated systems, to study separately the effect of structure and the effect of chemical nature of the functional groups in FAU type carbons.}
\label{fig:strategy}
\end{figure}
The functionalization will therefore have two effects at the same time: a chemical effect, related to specific interatomic interactions as oxygen and hydrogen are different from the carbon atoms (here, different charges and Lennard-Jones parameters); and a physical effect which corresponds to the modification of the carbon structure. To study these two effects separately, we adopted a strategy that consists in studying an additional system, ZTC-C, where the oxygen and hydrogen atoms are replaced by carbon atoms. This strategy is illustrated in Figure~\ref{fig:strategy}.

\section{Results and discussion}

\subsection{Influence of the functionalization on the quantity of adsorbed ions}

To be able to analyze the behaviour of a confined electrolyte, the first thing one has to do is to distinguish between species inside and outside the porous material. In this work, this distinction is simply based on the position of the ions. If the center of mass of an ion has a $z$ coordinate between $z_{min}$ and $z_{max}$, the $z$ coordinates of the outermost carbon atoms, the ion is considered inside the porous material. It is considered outside otherwise. Following this definition, it is possible to calculate the total number of adsorbed ions, referred to as the total pore population (TPP)~:
\begin{equation}
TPP = \frac{N_{anions}+N_{cations}}{m_{mat}}
\label{TPP}
\end{equation}
which is normalized either by the mass of the material ($m_{mat}$,  see equation~\ref{TPP}) or by the accessible volume (V$_{acc}$, determined using Poreblazer~\citep{Sarkisov11}).

\begin{figure} [ht!]
\begin{center}
\includegraphics[scale=0.29]{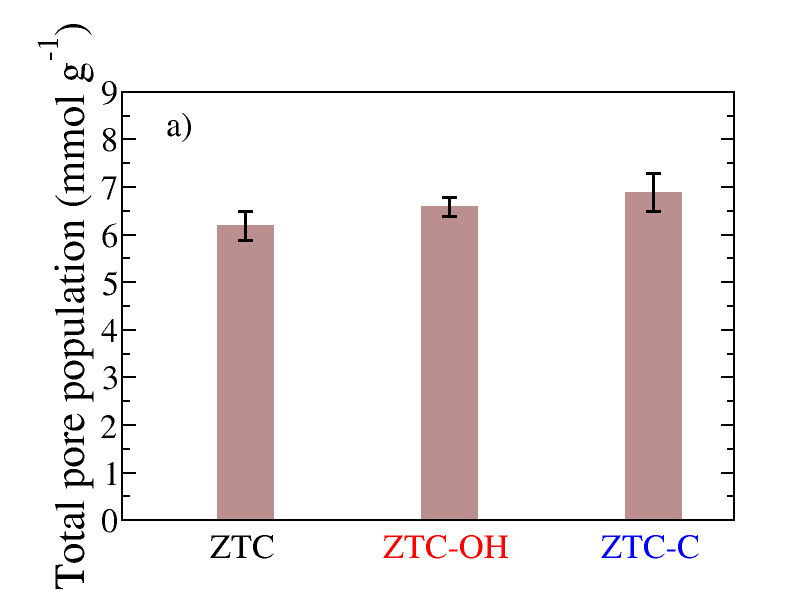}
\includegraphics[scale=0.29]{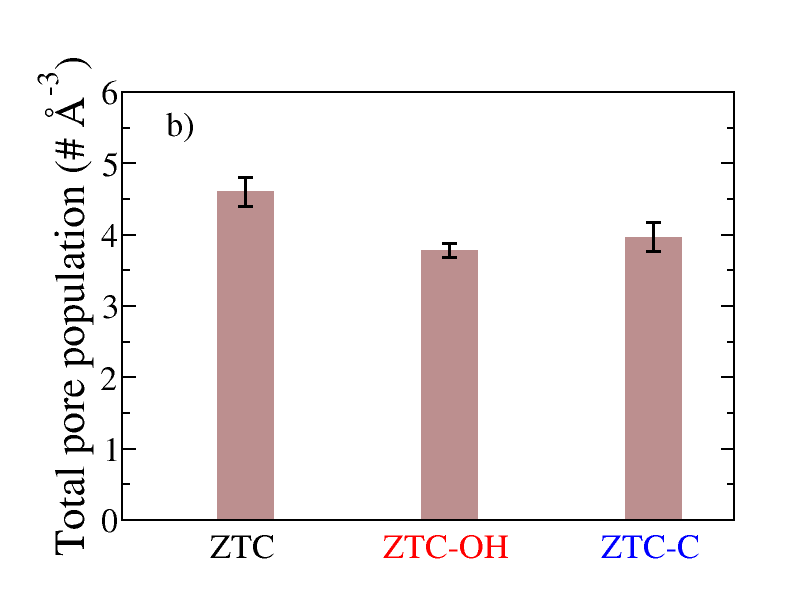}
\end{center}
\caption{Total pore populations in the three structures normalized by a) the mass of carbon material or b) by the accessible volume (\# designates a number of ions).}
\label{fig:TPP-realmasse}
\end{figure}

Figure~\ref{fig:TPP-realmasse} shows the TPP calculated for the three simulated systems. 
The mass values, the accessible volumes, the densities as well as the numbers of confined ions are provided in Table~\ref{table:info-FAU}. Indeed, although the overall volumes of the materials are similar, their characteristics and in particular porosities are different. It is important to note that the functionalized carbon, ZTC-OH, which was generated by \emph{ab initio} calculations, actually has a lower density than the initial ZTC carbon.
\begin{table}[ht!]
\centering
\begin{tabular}{|c|c|c|c|c|c|}
\hline
System & $m_{mat}$ (g~mol$^{-1}$) & $V_{acc}$ (Å$^3$) & $V_{acc}$ (cm$^{3}$~g$^{-1}$) & Density (g~cm$^{-3}$) & $N_{ions}$ \\
\hline
ZTC & 60 384 & 81 175 & 0.81 & 0.82 & 374 \\ 
\hline
ZTC-OH & 47 008 & 81 830 & 1.05 & 0.65 & 310 \\ 
\hline
ZTC-C & 46 848 & 81 352 & 1.05 & 0.66 & 324 \\ 
\hline
\end{tabular}
\caption{Masses ($m_{mat}$), geometrical accessible volumes ($V_{acc}$) and densities of the carbon materials, and number of adsorbed ions ($N_{ions} = N_{anions} + N_{cations}$) for the three simulated systems.}
\label{table:info-FAU}
\end{table}
There are not many experimental methods to assess quantitatively the amount of ions adsorbed in porous carbons. An experimental NMR study for the ionic liquids [Pyr$_{13}$][TFSI] and [EMI][TFSI] confined in an activated carbon (YP50F) provides TPP of 3.2~mmol~g$^{-1}$ and 3.6~mmol~g$^{-1}$ respectively~\citep{Forse15} (we double the values given in that article because of the different definitions of in-pore populations in moles of ionic liquid in the NMR work against the sum of anions and cations here). While being of the same order of magnitude, the values from the molecular simulations are significantly larger, between 6.2 and 6.9~mmol~g$^{-1}$. Forse~\emph{et~al.} mention a total pore volume of 0.71~cm$^{3}$~g$^{-1}$ for YP50F (measured through N$_2$ adsorption), smaller than the values of 0.81~cm$^{3}$~g$^{-1}$ and 1.05~cm$^{3}$~g$^{-1}$ reported here for the ZTC carbons which, in addition to the different ionic liquid nature, could explain this quite large difference. It is worth noting that experimental data reports larger pore volumes for ZTC carbons than for YP50F: Builes~\emph{et~al.}~\citep{Builes11} report a total pore volume of 1.69~cm$^{3}$~g$^{-1}$ and a microporous volume of 1.43~cm$^{3}$~g$^{-1}$. The pore volumes of the model atomistic carbons considered here are thus between the ones of YP50F and ZTC experimental materials.

Figure~\ref{fig:TPP-realmasse} shows that the TPP normalized by the mass of material increases slightly, +~0.4~mmol~g$^{-1}$, with the presence of functional groups. The standard deviations calculated between 0.2~mmol~g$^{-1}$ (ZTC-OH) and 0.4~mmol~g$^{ -1}$ (ZTC-C) suggests that these materials are in fact equivalent in terms of TPP normalized by the mass of the material.  For the TPP normalized by the accessible volume, a more significant decrease of 17.8\% is observed between ZTC and ZTC-OH. On the contrary, the difference between ZTC-OH and ZTC-C seems insignificant. These changes in the amount of confined ions for a similar geometrical accessible volume may be due to i) size reduction or blocking of certain pores due to the presence of functional groups, ii) the disorganization of the confined ionic liquid by the functional groups, or iii) a larger distance between the ions and the carbon surface.

\begin{figure} [ht!]
\begin{center}
\includegraphics[scale=0.34]{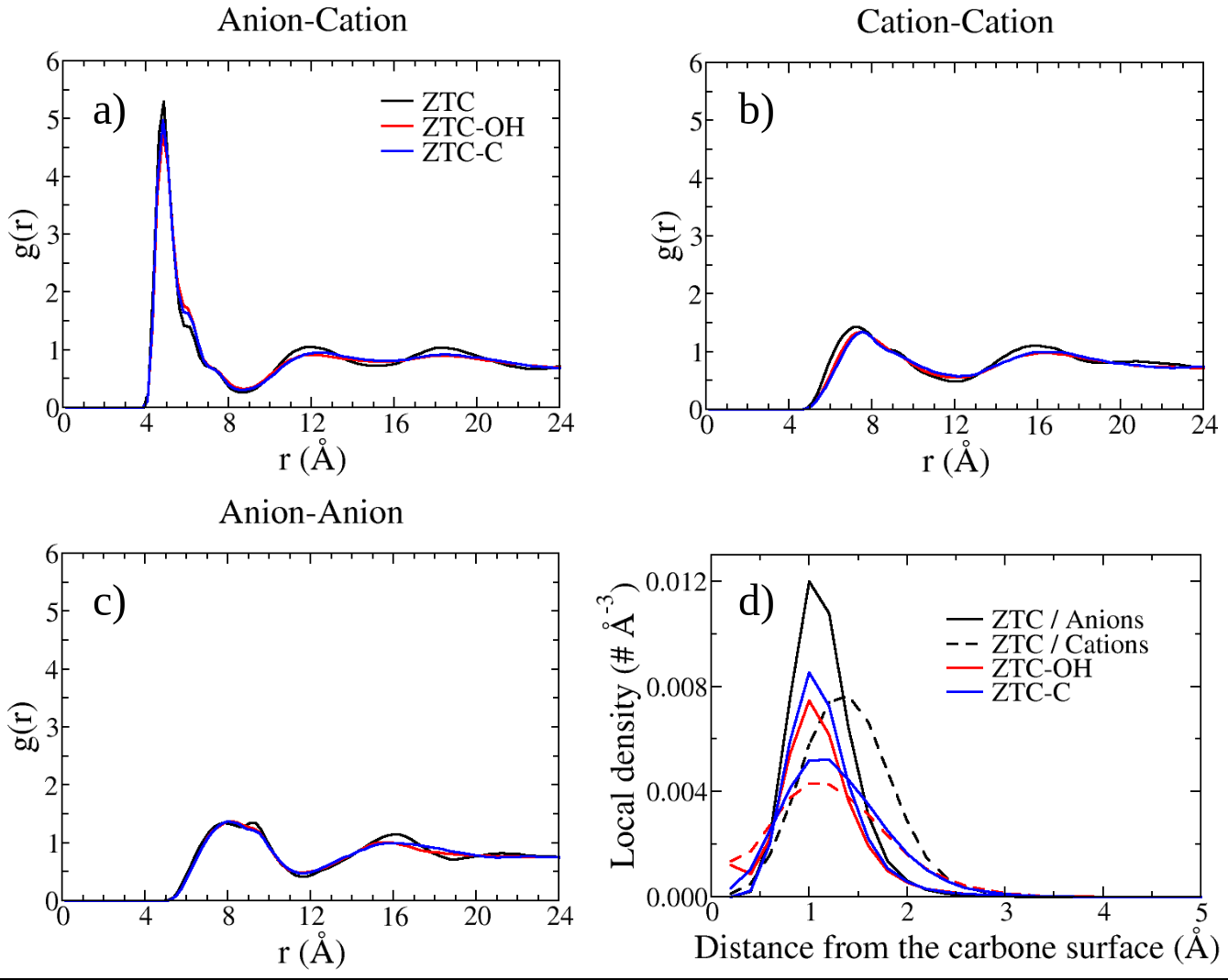}
\end{center}
\caption{Pair distribution functions (a, b, c) and local densities (d) for the centers of mass of the ions confined in the porous ZTC structures.}
\label{fig:PDF-FAUs}
\end{figure}

To test the hypothesis of a disorganization of the ionic liquid by the presence of functional groups / defects, we calculated pair distribution functions between the centers of mass of the ions. Figure~\ref{fig:PDF-FAUs} shows the pair distribution functions for the three simulated systems.
The presence of functional groups / defects has a limited impact on the local structure of the electrolyte. There is possibly a small reorientation of the cations visible with the slight increase in the shoulder at 6.0~\r{A} in the anion-cation distribution, and in the slight increase (+~3.4\%) in the most probable cation-cation distance, when going from ZTC to the ZTC-OH and ZTC-C. The second and third peaks of the pair distribution functions (at around 12.0~\r{A} and 18.4~\r{A} in the anion-cation distribution) also seem to be flattened for ZTC-OH and ZTC-C compared to the initial ZTC. Overall, it seems that the presence of functional groups / defects induces a small disorganization of the confined ionic liquid.

To test the hypothesis of a larger distance between the ions and the carbon surface, we calculated local densities following the methodology described by Merlet~\emph{et al.}.~\citep{Merlet12} which uses an argon probe with a diameter of 5.5~\r{A} to define the accessible surface and determines the local density as a function of the distance to the local surface. Figure~\ref{fig:PDF-FAUs}d shows the results of such analysis. For the anions, while there is a clear decrease in the intensity of the peak when going from ZTC to ZTC-OH and ZTC-C, directly related to the decrease in TPP observed in Figure~\ref{fig:TPP-realmasse}, its position is unchanged with the presence of functional groups / defects. For the cations, the decrease in the intensity of the peak is associated with a shift to smaller distances. Interestingly for both anions and cations, while the nature of the defects, O/H functional groups or C atoms, affects the height of the peaks, it does not influence their positions.

Following the results obtained through pair distribution functions and local ionic densities, it seems that the generation of small pores as seen in Figure~\ref{fig:PSD-FAU}, probably not accessible to the ions, is the main reason for the reduced TPP in ZTC-OH and ZTC-C compared to the intial ZTC.

\subsection{Diffusion coefficient and confinement}

\begin{figure} [ht!]
\begin{center}
\includegraphics[scale=0.4]{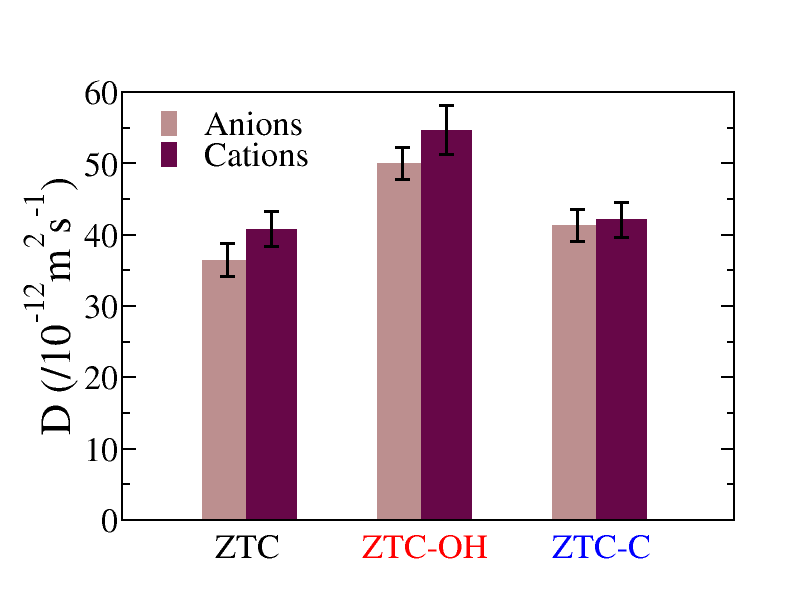}
\end{center}
\caption{Diffusion coefficients of the confined ions for the three simulated structures.}
\label{fig:Diff-FAUs}
\end{figure}

Figure~\ref{fig:Diff-FAUs} shows the diffusion coefficients for the ions confined in the three simulated materials. The diffusion coefficients for the ions in the bulk are around 200~$10^{-12}$~m$^2$~s$^{-1}$ for anions and 220~$10^{-12}$ m$^2$~s$^{-1}$ for cations.~\citep{Tokuda04,Roy10b,Lahrar20} As seen in previous studies,~\citep{Lahrar20,Kondrat14,Burt16,He16} the confinement leads to a large decrease in diffusion but, in most cases, less significant than what is observed experimentally.~\citep{Forse17} Regarding the effect of the functional groups, the diffusion coefficient of the confined ions increases by about 27\% for anions and 25\% for cations when going from ZTC to ZTC-OH. This behaviour is in agreement with experimental results showing an increase in diffusion with the addition of functional groups~\citep{Dyatkin16, Dyatkin14,Dyatkin18}. The diffusion coefficient of the confined ions increases only by about 11\% for the anions and 4\% for the cations when going from the ZTC to ZTC-C structure. This seems to indicate that the specific interactions between the oxygen and hydrogen atoms and the ions are important for diffusion, again in agreement with experimental results.~\citep{Dyatkin14,Dyatkin15}

\begin{figure} [ht!]
\begin{center}
\includegraphics[scale=0.29]{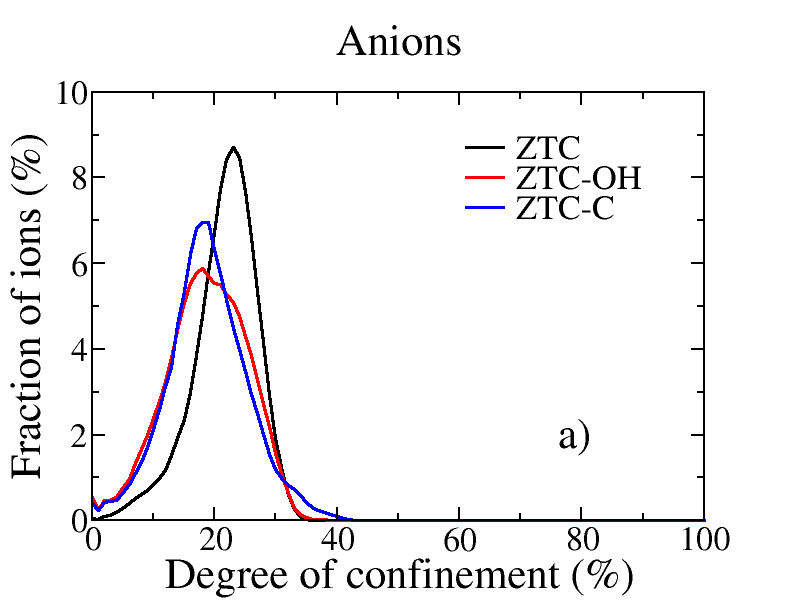}
\includegraphics[scale=0.29]{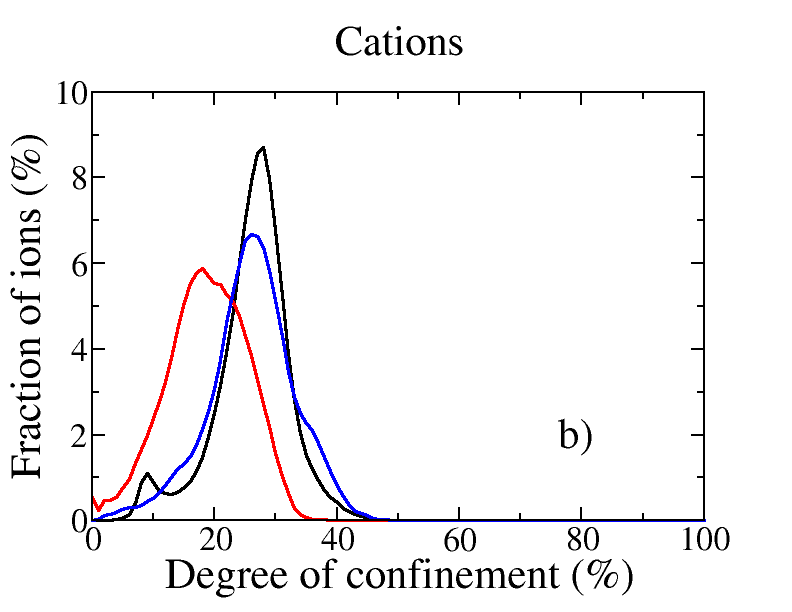}
\end{center}
\caption{Distribution of degrees of confinement of confined anions and cations in the porous ZTC carbons.}
\label{fig:DoC}
\end{figure}

To understand these observations, we calculated the degrees of confinement of the ions in the three porous structures. Indeed, a previous study has shown that there is a correlation between confinement and diffusion.~\citep{Lahrar20}. The degree of confinement (DoC), as defined by Merlet~\emph{et~al.}~\citep{Merlet13d}, is the percentage of the solid angle around the ion which is occupied by the carbon atoms, normalized by the maximal value taken by this quantity. The DoC depends both on the number of carbons surrounding an ion and on each ion-carbon distance. Figure~\ref{fig:DoC} gives the distributions of anions and cations in different degrees of confinement.
This figure shows that, in most cases, the ions become less confined when surface functional groups / defects are present which is consistent with an increase in diffusion coefficients. The DoCs of the anions in ZTC-OH and ZTC-C structures are similar and shifted to lower confinement (peak maximum at $\sim$19\%) compared to ZTC (peak maximum at $\sim$23\%). It is interesting to note that since local densities in Figure~\ref{fig:PDF-FAUs}d show that the anion-carbon distance does not change, this indicates that anions are occupying environments where they are surrounded by less carbon atoms. For the cations, the DoC curves for ZTC and ZTC-C have a similar main peak but i) the peak at $\sim$10\% in ZTC is not present in ZTC-C, a shoulder at 14\% is present instead, and ii) a shoulder at $\sim$ 36\% is seen for ZTC-C. The curve corresponding to ZTC-OH shows a more dramatic variation of the position of the main peak from $\sim$27-28\% to $\sim$19\% (as for the anions). In this case, the average cation-carbon distance was shown to vary with the functionalization. Anions are spherical with a single charge while cations consist in three sites carrying different partial charges. The cations are therefore subject to reorientation, which is not possible with anions. In ZTC-C, the smaller cation-carbon distance compared to ZTC (see Figure~\ref{fig:PDF-FAUs}) could be associated with the increase of DoC (from $\sim$10\% to $\sim$14\% and from $\sim$27-28\% to $\sim$36\%). In ZTC-OH, the surface functional groups have partial charges which will induce electrostatic interactions with the ions and probably affect the orientation of the cations. In this case, the environment of the cations is more largely affected, maybe with less carbon atoms in their first coordination shell, and the smaller cation-carbon distance is not associated with an increase in DoC. This can explain the different behaviour observed for ZTC-OH compared to ZTC and ZTC-C. Overall, for the anions, the modification of the structure seems to play a more important role than the chemical nature of the groups while it is the reverse behavior for the cations.

Overall, the molecular simulations conducted here on a neat ionic liquid in contact with several three-dimensional porous carbon structures confirm previous results~\citep{Dyatkin15,Dyatkin16,Kerisit14,DeYoung14,Wang18,Schweiser19} showing that specific interactions between the ions and the carbon material affect the structural and dynamical properties of the confined ions in different ways. Steric and chemical variations associated with the presence of functional groups, and having possibly contradictory effects, make it difficult to predict the influence of the presence of functional groups on the properties of the confined ionic liquid. The ion geometry and charge distribution potentially affecting the steric and specific interactions, it would be interesting to explore the impact of using an all-atom model to represent the ionic liquid in future works.

\section{Conclusions}

In this work, we have investigated the influence of the presence of surface functional groups (-O and -H) on the properties of a confined ionic liquid. We carried out simulations with ``Zeolite Templated Carbons'' having an ordered structure, with and without functional groups. In addition to the initial ZTC structure, without any defects or functional groups, and the ZTC-OH structure, functionalized using DFT simulations of the oxidation, we have studied a structure where the oxygens and hydrogens are replaced by carbon atoms in order to dissociate steric effects and more specific effects. The molecular simulations show that the functionalization affects both the quantities of adsorbed ions and diffusion coefficients without large variations in the local structure of the ionic liquid. In particular, ions are faster for functionalized carbons, in agreement with experimental results, an effect which could be rationalised by characterizing the confinement of the ions. For the systems studied here, the quantities of adsorbed ions seem to be more affected by steric interactions but for other properties, like diffusion coefficients and degrees of confinement, the relative importance of steric and specific interactions depends on the ion type. In the future, it would be interesting to explore the influence of representing the ionic liquid with an all-atom model and the impact of allowing a certain flexibility for the carbon structure.

\section*{Data availability}

The data corresponding to the plots reported in this paper, as well as example input files for LAMMPS, are available in the Zenodo repository with identifier 10.5281/zenodo.4446690.

\begin{acknowledgement}
This project has received funding from the European Research Council (ERC) under the European Union's Horizon 2020 research and innovation program (grant agreement no. 714581). This work was granted access to the HPC resources of CALMIP supercomputing center under the allocation P17037.
\end{acknowledgement}



\providecommand{\latin}[1]{#1}
\providecommand*\mcitethebibliography{\thebibliography}
\csname @ifundefined\endcsname{endmcitethebibliography}
  {\let\endmcitethebibliography\endthebibliography}{}

\end{document}